\documentclass[twocolumn,twoside]{article}
\usepackage{graphics}
\usepackage[T1,T2A]{fontenc}  
\usepackage[utf8]{inputenc}
\usepackage[ukrainian,english]{babel}

\newcommand{\hb}{\\ \hspace*{2ex}}

\begin{document}
\title{X-RAY EMISSION OF ICRF SOURCES}
\author{V.\,V. Voitsekhovskiy$^{1}$, A.\,V. Tugay$^{1}$, V.\,V. Tkachuk$^{1}$, S.\,Yu. Shevchenko$^{2}$ \\[2mm] 
\begin{tabular}{l}
 $^1$ Taras Shevchenko National University of Kyiv,\hb
 Kyiv, Ukraine,  {\em tugay.anatoliy@gmail.com}\\
 $^2$ Schmalhausen Institute of Zoology,\hb
 Kyiv, Ukraine,  {\em astromott@gmail.com}\\[2mm]
\end{tabular}
}
\date{}
\maketitle
%
%
ABSTRACT.
Considering increasing requirements to the coordinates measurement precision by the end of XX century International Astronomical Union commenced implementation of the  new astrometric system  ICRF (International Celestial Reference Frame). This quasi-inertial reference frame system centered in the barycenter of the Solar System and  has axes defined by the positions of distant extragalactic sources – frames. Unlike equatorial system ICRF has no shortcomings of the coordinates identification due to the Earth axis precession, stellar proper motions and other factors.
Extragalactic frames of the ICRF system are mostly quasars, radio galaxies, blazars and Seyfert galaxies i.e. different types of the active galaxy nuclei (AGN).  Active galaxy nuclei are characterized by processes with significant. Such processes quite often are followed by X-ray emission generation.   The purpose of this work is to consider X-ray emission of ICRF sources and features of their possible proper motions. Among 295 selected reference frames of the system we identified 54 X-ray  sources which were observed by space observatory XMM Newton and noticed rapid variability of the blazars 2E 2673 (W Com) and 2E 1802 which enables to conclude that they have some very active processes in the sources centers. 
With regards to the future more detailed analysis we beleive that evidences of the objects proper motion could be found in their spectra. 
Based on the constructed luminosity and spectral graphs we could conclude that apart from above mentioned AGNs rest 52 objects do not show veriability and special attention should be paid to blazars within ICRF system development and use. Major part of the X-ray sources between the reference frames are stable. \\[1mm]
\\[2mm]
РЕНТГЕНІВСЬКЕ ВИПРОМІНЮВАННЯ ДЖЕРЕЛЕ ICRF.
Зважаючи на зростаючі вимоги до точності вимірювання координат, наприкінці XX століття Міжнародний Астрономічний союз запровадив нову астрометричну  систему ICRF (International Celestial Reference Frame). Ця квазіінерціальна система координат має центр в барицентрі Сонячної системи та її вісі визначаються  положенням віддалених позагалактичних джерел - фреймів. На відміну від екваторіальної системи,  ICRF позбавлена недоліків визначення координат, що повязані з прецесією, власним рухом зірок та іншими факторами.    
Позагалактичні джерела системи ICRF являють собою, як правило, квазари, радіогалактики, блазари і сейфертівські галактики, тобто різні типи активних ядер галактик. Активні ядра галактик (АЯГ) характеризуються процесами з значними швидкими рухами. Такі процеси часто супроводжуються виникненням рентгенівського випромінення. Метою даної роботи є вивчення рентгенівського випромінювання джерел ICRF та особливості їх можливого власного руху. Серед 295 визначених опорних фреймів системи ми знайшли 54 рентгенівських джерела, які спостерігалися космічною обсерваторією XMM-Newton та встановили, що швидку мінливість спектру мають блазари 2E 2673 (W Com) та 2E 1802, що дозволяє зробити висновок про дуже активні процеси в центрі джерел. З огляду на проведення більш повного аналізу ми вважаємо, що можуть знайтися свідоцтва власних рухів в їх спектрах. Базуючись на побудованих кривих блиску та спектрах можна зазначити, що окрім вищенаведених активних ядер галактик 52 об'єкти не демонструють змінності.   Ми прийшли до висновку, що в рамках розробки та використання системи  ICRF необхідно приділяти особливу увагу блазарам.  Більшість рентгенівьских джерел в опорних фреймах системи є стабільними.\\[1mm]
{\bf Ключові слова}: ICRF, система опорних координат, галактики: активні, рентгенівське випромінення: галактики.
\\[2mm]
%
%
{\bf 1. Introduction}\\[1mm]

ICRF is a frame of reference defined by the positions of extragalactic sources. The most appropriate extragalactic sources should be very luminous so it is obvious that many of them are active galactic nuclei (AGNs). They always has some active processes and rapid motion in the center. 

Active galactic nuclei are complex phenomena. 
At the heart of an AGN there is a relativistic accretion disk around the spinning supermassive black hole.
In the center of the galaxy the emissions of relativistic particles occurs as narrow and uniform brightness jets.
Curious that, their observation may indicate that the center of image is moving with speed over the speed of light (Schneider, 2006).
The explanation of this effect is that we can only observe the movement of a projection on area perpendicular to the field of vision.
The effect of "superluminal" expansion is observed in jets moving toward the observer at a small angle to the line of sight. 
Blazars has the smallest value of this angle among other types of AGNs, so we considered to check spectra for available BL Lacs in ICRF to identify possible motion. 

The accuracy of ICRF is 40 microarcsec\footnote{http://hpiers.obspm.fr/icrs-pc/, 'ICRF2' chapter}. AGN with typical redshift z=1 may have relativistic radiojet with knots passing through this angle in few years. 
Such relativistic motion in the AGNs could be detected by astrometric methods in the nearest future. Some signs of these processes can be checked in X-ray band. X-ray emission is generated in the very center of AGN and is influenced by the fastest motion of the matter. 
The aim of this work is to consider X-ray emission of ICRF sources and to look for signs of their possible proper motion. 
For this purpose we used 3XMM-DR4 catalog (Rosen et al., 2015) of X-ray sources based on observational data from modern space observatory XMM-Newton. 
We checked light curves of all X-ray ICRF sources and made certain conclusions about their stability.\\[2mm]

{\bf 2. Identification and analysis of X-ray ICRF sources}\\[1mm]

Initial phase of our work was to find ICRF sources in 3XMM-DR4 catalog. Current version of ICRF (ICRF2) composed by 3414 sources\footnote{http://rorf.usno.navy.mil/ICRF2/} and 3XMM-DR4 catalog contains 372728 individual sources. 
Using X-ray and optical galaxies from cross-identification sample (Tugay, 2012) we found that 54 of ICRF sources are listed in the 3XMM-DR4 catalog. 
33 sources of the sample have only one XMM observation and the rest 21 sources have two observations each.
General parameters of these sources are presented in the Table 1 below.
We reviewed X-ray light curves of these sources using LEDAS database and found that most of them are constant.
The deviation of mean X-ray flux of 52 objects during the whole observation does not exceed the half of rms scatter. 
The example of constant X-ray light curve is presented at Fig. 1. 
However, we found X-ray variability of 2E 1802 and 2E 2673 (W Com) blazars during XMM observations dated 04.04.2004 and 22.06.2002 respectively (Fig. 2-3). Such variability can not be explained by soft proton solar flares which were observed for a number of other ICRF sources. So these two blazars are typical examples of the most active galactic nuclei with rapid and powerful processes in the center. Other blazars may also have variability out of XMM observation time.
We analyzed spectra of all blazars from our sample to check any possible special features and to reveal some additional unusual objects.
We applied standard XMM SAS software to fit two-parametric power law model of X-ray continuum with hydrogen absorption at low energies. The results of spectral fitting are given in the Table 2 and some examples of spectra are shown on Figs. 4-6.
All blazar spectra have a good approximation by our simple model. We assume that if the spectrum of any blazar is not fitted by such simple model then there is a reason to expect some complex structure of the source and possible variability. Otherwise, we consider the blazar as a typical source and appropriate for ICRF usage.

\begin{figure}[h] 
\resizebox{1.0\hsize}{!}
{\includegraphics{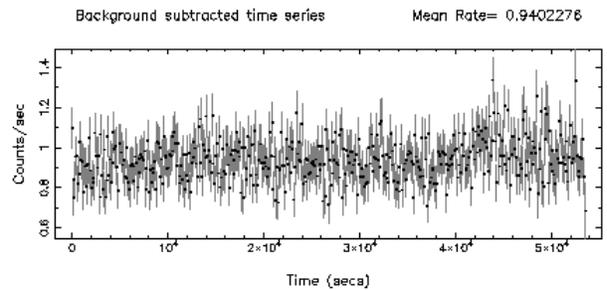}}
\label{hh}
\caption{Light curve of OJ287. Example of stable X-ray source.}
\end{figure}

\begin{figure}[h] 
\resizebox{1.0\hsize}{!}
{\includegraphics{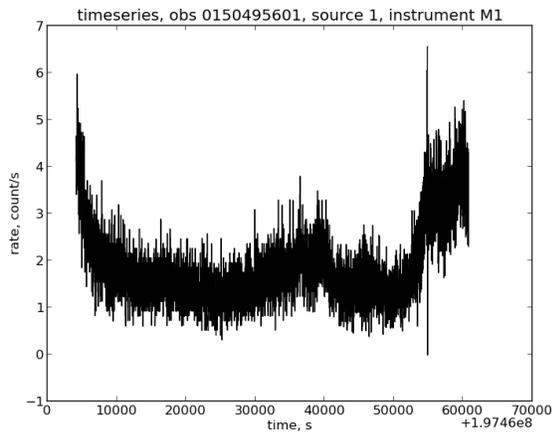}}
\label{hh}
\caption{Light curve of 2E 1802 from MOS1 camera. The time at Fig. 2 and 3 is measured in seconds from 01.01.1998.}
\end{figure}

\begin{figure}[h] 
\resizebox{1.0\hsize}{!}
{\includegraphics{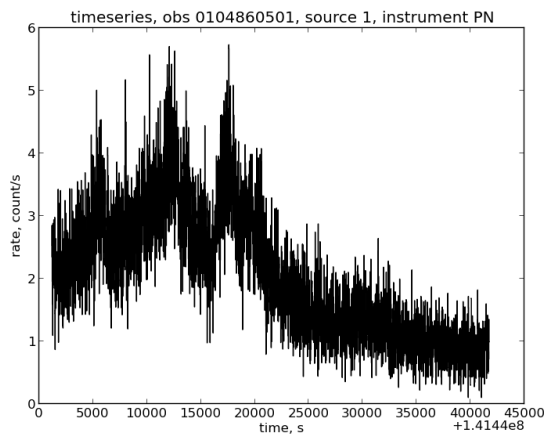}}
\label{hh}
\caption{Light curve of W Com from PN camera.}
\end{figure}

\begin{figure}[h] 
\resizebox{1.0\hsize}{!}
{\includegraphics{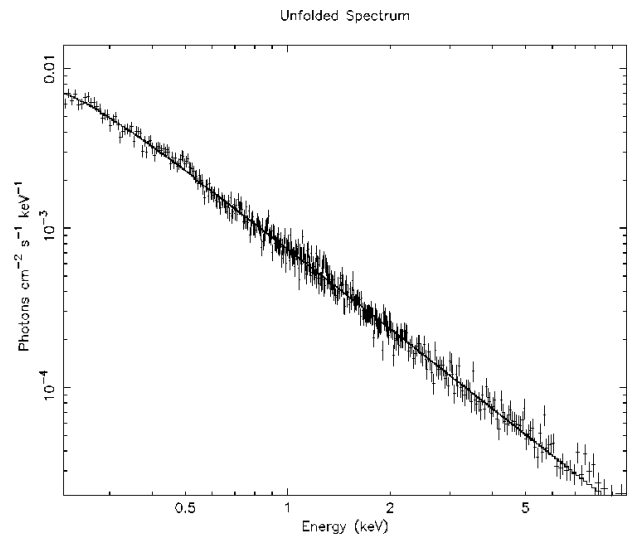}}
\label{hh}
\caption{Spectrum of 2E 2979. Example of unabsorbed source.}
\end{figure}

Spectral parameters are constant in all available XMM observations. So we have found out that X-ray spectra makes no reasons to expect any image motion of these sources in the nearest future. We conclude that current state of spectral analysis doesn't allow to predict a correlation between the spectra parameters and variability.\\[2mm]

%
%

{\bf 3. Results and conclusions}\\[1mm]

 In the result of our analysis of light curves and spectral characteristics of ICRF blazars in X-ray band we found rapid variability for 2E 2673 (W Com) and 2E 1802 but no special features in the spectra. 

Assuming that the variability of a source in any band should be connected with possible motion of the source (and the center of its image)
 we might predict that stable source should not have significant motion. 

So we conclude (for the moment) that the most of ICRF sources are stable and appropriate as defined sources for astrometric reference systems. 
Keeping in mind all possible extreme physical processes related to active galactic nuclei, the issue of possible image motion of ICRF sources remains persistent. W Com and 2E 1802 should still be considered as 'suspicious' ICRF sources. 
Their X-ray variability reveals fast processes in the very center of AGN that could cause detection of the image shift in the future.
Future radiotelescopic studies of astrometric defined sources should review more thoroughly not only these two objects but any blazars, since the luminosity variability or even a motion could be detected.

%
%

%
%
{\it Acknowledgements.} The authors are thankful to Vasyl Choliy for motivation to such interesting and actual task. 
\\[3mm]
%
%
{\bf References\\[2mm]}
Baumgartner~W.\,H., Tueller~J., Markwardt~C. et. al. 2012, {\it ApJ SS}, {\bf 207}, 19 \\
Cusumano~G., LaParola~V., Segreto~A. et al. 2010, {\it A\& A}, {\bf 524}, A64 \\
Donato~D., Ghisellini~G., Tagliaferri~G. et. al. 2001, {\it A\& A}, {\bf 375}, 739 \\
Grandi~P. \& Malaguti~G. 2006, {\it ApJ,} {\bf 642}, 113 \\
Ma C., Arias E.F., Eubanks T.F. et al. 1998, {\it ApJ,} {\bf 116}, 516 \\
Piconcelli~E., Cappi~M., Bassani~L. et. al. 2002, {\it A\& A}, {\bf 394}, 835 \\
Reeves~J.\,N. \& Turner~M.\,L. 2000, {\it Mon. Not. R. Astron. Soc.,} {\bf 316}, 234 \\
Rosen S., Watson M., Pye J. et al. 2015, {\it ADASS XXIV Proc.,} 319. \\
Schneider P. 2006. 'Extragalactic Astronomy and Cosmology', Springer, Berlin Heidelberg New York \\
Tugay A. 2012. {\it Odessa Astron. Publ.,} {\bf 25}, 142 \\
Winter~L., Muschoztky~R., Reynolds~C.\,R. et. al. 2010, {\it AIP Conference Proceedings,} {\bf 369}, 1248 \\
\vfill
%

%
%

\begin{table}
 \centering
 \caption{List of X-ray ICRF sources.}\label{tab1}
 \vspace*{1ex}
 \begin{tabular}{lllll}
  \hline
  N & XMM ID & SIMBAD type & Name & z \\
  \hline
  1.  & J001031.0+105829 & Seyfert 1    & Mrk 1501 & 0.090 \\
  2.  & J003824.8+413706 & Quasar       & 5C 3.50 & 1.353 \\
  3.  & J005748.8+302108 & Galaxy       & IAU 0057+3021 & 0.016 \\
  4.  & J012642.7+255901 & Quasar       & 4C 25.05 & 2.358 \\
  5.  & J014922.3+055553 & Quasar       & QSO B0146+056 & 2.347 \\
  6.  & J015002.6-072548 & Seyfert 2    & IAU 0150-0725 & 0.017 \\
  7.  & J022239.6+430207 & BL Lac       & 3C 66A & 0.340 \\
  8.  & J023838.9+163659 & BL Lac       & 2E 618 & 0.940 \\
  9.  & J024008.1-230915 & Quasar       & 2E 638 & 2.225 \\
  10. & J024457.6+622806 & Seyfert 1    & 2E 653 & 0.044 \\
  11. & J031155.2-765150 & BL Lac       & 2E 746 & 0.223 \\
  12. & J031301.9+412001 & Seyfert 1    & QSO B0309+411 & 0.136 \\
  13. & J040748.4-121136 & Seyfert 1    & 2E 938 & 0.572 \\
  14. & J044017.1-433308 & Quasar       & 2E 1127 & 2.852 \\
  15. & J052257.9-362730 & BL Lac       & 2E 1263 & 0.055 \\
  16. & J053056.4+133155 & Quasar       & 2E 1289 & 2.070 \\
  17. & J053954.2-283955 & Quasar       & 2E 1496 & 3.103 \\
  18. & J072153.4+712036 & BL Lac       & 2E 1802 & 0.300 \\
  19. & J084124.3+705342 & Quasar       & 4C 71.0 & 2.172 \\
  20. & J085448.8+200630 & BL Lac       & OJ 287 & 0.306 \\
  21. & J095456.8+174331 & Quasar       & QSO B0952+179 & 1.475 \\
  22. & J095524.7+690113 & Super Nova   & SN 1993J & 0.0001 \\
  23. & J104117.1+061016 & Quasar       & 2E 2303 & 1.270 \\
  24. & J105829.6+013358 & BL Lac       & 4C 01.28 & 0.185 \\
  25. & J112027.8+142054 & LINER        & 4C 14.41 & 0.362 \\
  26. & J113007.0-144927 & Quasar       & 2E 2471 & 1.189 \\
  27. & J121923.2+054929 & LINER        & NGC 4261 & 0.007 \\
  28. & J122006.8+291650 & LINER        & NGC 4278 & 0.002 \\
  29. & J122131.6+281358 & BL Lac       & 2E 2673 & 0.102 \\
  30. & J122222.5+041315 & Quasar       & 4C 04.42 & 0.965 \\
  31. & J122906.6+020308 & BL Lac       & 3C 273 & 0.173 \\
  32. & J123959.4-113722 & LINER        & M 104 & 0.003 \\
  33. & J124646.8-254749 & Quasar       & QSO B1244-25 &  0.638 \\
  34. & J125359.5-405930 & Quasar       & QSO B1251-407 & 4.464 \\
  35. & J125611.1-054721 & Quasar       & 3C 279 & 0.536 \\
  36. & J130533.0-103319 & Seyfert 1    & 2E 2966 & 0.278 \\
  37. & J131028.6+322043 & BL Lac       & 2E 2979 & 0.997 \\
  38. & J132527.6-430108 & Seyfert 2    & CENTAURUS A & 0.001 \\
  39. & J132616.5+315409 & Radio Galaxy & 4C 32.44 & 0.370 \\
  40. & J140700.3+282714 & BL Lac       & Mrk 668 & 0.076 \\
  41. & J140856.4-075226 & Quasar       & QSO B1406-076 & 1.493 \\
  42. & J143023.7+420436 & Quasar       & 7C 1428+4218 & 4.705 \\
  43. & J151002.9+570243 & Quasar       & 1E 1508.7+5714 & 3.880\\
  44. & J160913.3+264129 & Radio Galaxy & PKS 1607+268 & 0.473 \\
  45. & J165352.2+394536 & BL Lac       & Mrk 501 & 0.033 \\
  46. & J170934.3-172853 & Quasar       & IAU 1709-1728 & 0.560 \\
  47. & J184208.9+794617 & Seyfert 1    & 2E 4136 & 0.056 \\
  48. & J201114.2-064403 & Radio Galaxy & PKS 2008-068 & 0.547 \\
  49. & J212912.1-153841 & Quasar       & 2E 4479 & 3.268 \\
  50. & J213032.8+050217 & Quasar       & IAU 2130+0502 & 0.990 \\
  51. & J215155.5-302753 & Quasar       & QSO B2149-307 & 2.344 \\
  52. & J220314.9+314538 & BL Lac       & 4C 31.63 & 0.295 \\ 
  53. & J225357.7+160853 & Quasar       & 3C 454.3 & 0.859 \\
  54. & J235509.4+495008 & Seyfert 2    & IAU 2355+4950 & 0.237 \\
  \hline 
 \end{tabular}
\end{table}

\clearpage

\begin{figure}[h] 
\resizebox{1.0\hsize}{!}
{\includegraphics{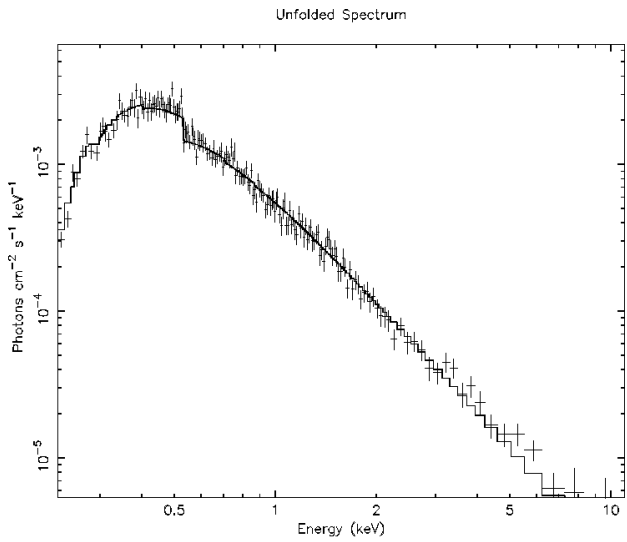}}
\label{hh}
\caption{Spectrum of 4C 01.28. Example of absorbed source.}
\end{figure}

\begin{table}
 \centering
 \caption{Spectral properties of ICRF BL Lacs. Normalisation unit is $10^{-5} cm^{-2}s^{-1}keV^{-1}$}\label{tab2}
 \vspace*{1ex}
 \begin{tabular}{lllll}
  \hline
  $Object$ & $n_H, 10^{20} cm^{-2} $ & Normalisation & $Photon Index$ & $\chi ^2$/d.o.f. \\
  \hline
  3C66A & 7.47 $\pm$ 0.49 & 66.2 $\pm$ 2.9 & 2.5 $\pm$ 0.04 & 1.0698 / 173 \\
  2E618 (Grandi, 2006) & 1.34 $\pm$ 0.03 &  & 1.55 $\pm$ 0.02 & 0.857 / 109 \\
  2E1263 (Winter, 2010; Cusumano, 2010) & 6 $\pm$ 1 &  & 2.14 $\pm$ 0.01 & 3.65 / 336 \\
  2E1802 (Baumgartner, 2012) & 3.81 $\pm$ 0.42 & & 2.7 $\pm$ 0.01 & 9.22 / 89 \\
  OJ287 & 20 $\pm$ 4 & 44.6 $\pm$ 1.1 & 1.75 $\pm$ 0.01 & 3.046 / 324 \\
  4C01.28 & 1.89 $\pm$ 0.23 & 55.8 $\pm$ 1.5 & 1.814 $\pm$ 0.018 & 1.0073 / 305 \\
  2E2673 (Piconcelli, 2002) & 0.64 $\pm$ 0.37 & & 1.32 $\pm$ 0.09 & 1.73 / 196 \\
  3C273 (Reeves, 2000; Donato, 2001) & 1.79 $\pm$ 0.2 & & 1.55 $\pm$ 0.02 & 0.857 / 1093 \\
  2E2979 & 0.432 $\pm$ 1.28 & 74.5 $\pm$ 1.3 & 1.668 $\pm$ 0.011 & 1.0655 / 399 \\
  Mrk668 & <0.01 & 2.2 $\pm$ 0.66 & 0.98 $\pm$ 0.13 & 2.905 / 8 \\
  Mrk501 & 2.70 $\pm$ 0.07 & 991 $\pm$ 40 & 2.49 $\pm$ 0.06 & 1.7800 / 316 \\
  \hline 
 \end{tabular}
\end{table}

\begin{figure}[h] 
\resizebox{1.0\hsize}{!}
{\includegraphics{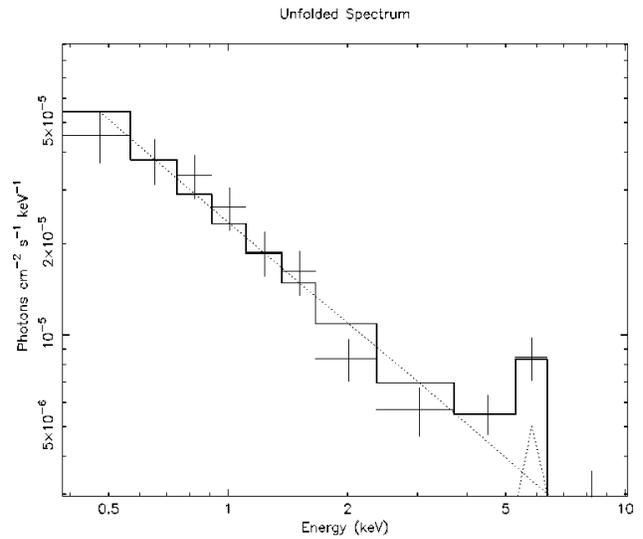}}
\label{hh}
\caption{Spectrum of Mrk 668. Example of faint source.}
\end{figure}

\end{document}